\documentclass[aip,jcp,11pt,reprint,floatfix,superscriptaddress,urlname,longbibliography]{revtex4-1}

\usepackage[utf8]{inputenc}

\usepackage{booktabs}
\usepackage{amsmath}
\usepackage{color}
\usepackage{comment}
\usepackage{algorithm}
\usepackage{graphicx}
\usepackage{xcolor}

\usepackage{hyperref}
\hypersetup{
    colorlinks,
    linkcolor={red!50!black},
    citecolor={red!70!black},
    urlcolor={red!80!black}
}

\newcommand{\angstrom}{\textup{\AA}~}
\DeclareMathAlphabet      {\mathbfit}{OML}{cmm}{b}{it}

\newcommand{\fnm}{\footnotemark}
\newcommand{\fnt}{\footnotetext}
\newcommand{\mc}{\multicolumn}
\usepackage{amssymb}
\newcommand{\T}[1]{#1^{\intercal}}
\usepackage[version=4]{mhchem}
\usepackage{physics}
\usepackage{siunitx}

\newcommand{\bO}{\boldsymbol{0}}

\newcommand{\bA}{\boldsymbol{A}}
\newcommand{\bB}{\boldsymbol{B}}

\newcommand{\bX}{\boldsymbol{X}}
\newcommand{\bY}{\boldsymbol{Y}}
\newcommand{\bV}{\boldsymbol{V}}
\newcommand{\bH}{\boldsymbol{H}}
\newcommand{\bG}{\boldsymbol{G}}
\newcommand{\bR}{\boldsymbol{R}}

\newcommand{\bU}{\boldsymbol{U}}
\newcommand{\bt}{\boldsymbol{t}}

\newcommand{\cE}{\mathcal{E}}
\newcommand{\cL}{\mathcal{L}}

\newcommand{\eps}{\epsilon}

\newcommand{\ii}{\text{i}}
\newcommand{\tI}{{\beta}}
\newcommand{\tJ}{{\gamma}}

\newcommand{\HF}{\text{HF}}
\newcommand{\QP}{\text{QP}}

\newcommand{\RPA}{\text{RPA}}

\newcommand{\TDA}{\text{TDA}}

\begin{document}

% title
\title{Fully Analytic Nuclear Gradients for the Bethe--Salpeter Equation}

% authors & affiliations
\author{Johannes T\"olle$^*$}
	\email{jojotoel@gmail.com}
	\affiliation{ Department of Chemistry, University of Hamburg, 22761 Hamburg, Germany;  The Hamburg Centre for Ultrafast Imaging (CUI), Hamburg 22761, Germany}
\author{Marios-Petros Kitsaras}
	\affiliation{Laboratoire de Chimie et Physique Quantiques (UMR 5626), Université de Toulouse, CNRS, Toulouse, France}
\author{Pierre-Fran\c{c}ois Loos}
	\affiliation{Laboratoire de Chimie et Physique Quantiques (UMR 5626), Université de Toulouse, CNRS, Toulouse, France}

% abstract 
\begin{abstract}
\textbf{Abstract:} The Bethe--Salpeter equation (BSE) formalism, combined with the $GW$ approximation for ionization energies and electron affinities, is emerging as an efficient and accurate method for predicting optical excitations in molecules.
In this letter, we present the first derivation and implementation of fully analytic nuclear gradients for the BSE@$G_0W_0$ method.
Building on recent developments for $G_0W_0$ nuclear gradients, we derive analytic nuclear gradients for several BSE@$G_0W_0$ variants.
We validate our implementation against numerical gradients and compare excited-state geometries and adiabatic excitation energies obtained from different BSE@$G_0W_0$ variants with those from state-of-the-art wavefunction methods.
\\
\begin{center}
	\boxed{\includegraphics[width=5cm, height=5cm]{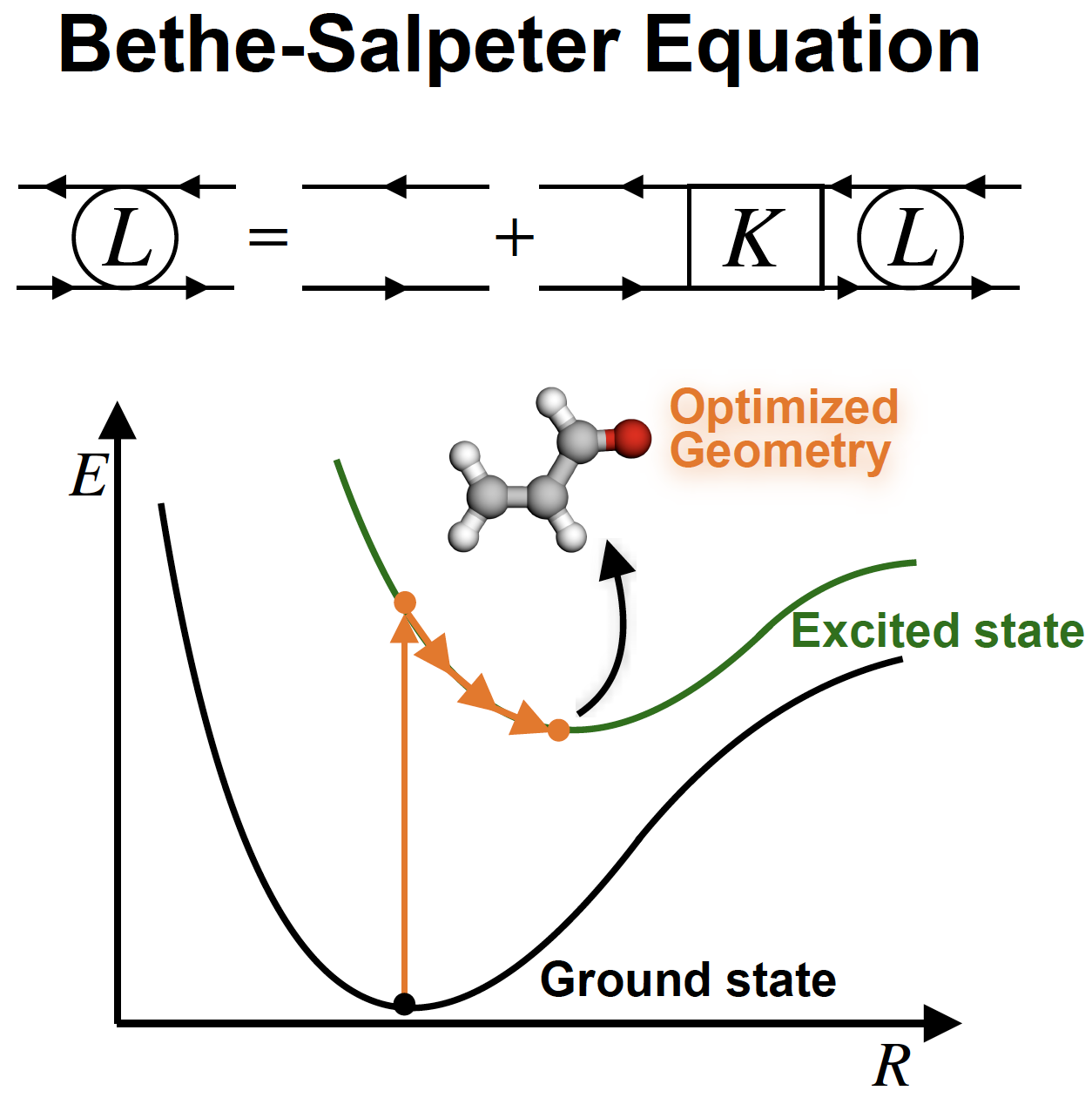}}
\end{center}
\end{abstract}

\maketitle 

% introduction
%\section{Introduction}

The Bethe--Salpeter equation (BSE) formalism \cite{salpeter1951relativistic,strinati1988application} combined with the $GW$ approximation \cite{hedin1965new,martin2016interacting,reining2018gw,golze2019gw,marie2023gw} of many-body perturbation theory (BSE@$GW$) was initially developed and widely adopted for computing vertical (neutral) excitation energies in solids. \cite{hanke1979many,strinati1982aspect,strinati1982shift,strinati1984effects,onida2002electronic,albrecht1997ab,albrecht1998ab,rohlfing1998excitonic,rohlfing1998electron,rohlfing2000electron,rohlfing1999optical}
More recently, BSE@$GW$ is steadily gaining popularity for the study of molecular excited states, \cite{grossman2001high,tiago2005first,rocca2010ab,blase2011charge,baumeier2012excited,duchemin2012short,bruneval2015systematic,leng2015excitons,hirose2015all,krause2017implementation,blase2018bethe,blase2020bethe,liu2020all,mckeon2022optimally} offering a favorable balance between computational cost and accuracy, \cite{onida2002electronic,blase2018bethe,blase2020bethe} often rivaling more computationally demanding wavefunction-based approaches. \cite{jacquemin2017bethe,blase2018bethe,loos2020quest}

In Ref.~\onlinecite{loos2020quest}, the accuracy of the BSE@$GW$ method for the singlet states in ``typical organic'' molecules was estimated to be around $0.1$--$0.3$ eV, comparable to approximate second-order coupled-cluster (CC) method (CC2), \cite{christiansen1995second,hattig2000excitation} or sometimes even full CC with single and double excitations (CCSD). \cite{koch1990excitation,stanton1993equation}
Larger errors are typically observed for triplet states \cite{jacquemin2017benchmark,rangel2017assessment,knysh2024reference} and double excitations. \cite{bintrim2022full} 
The latter is also a well-documented limitation of wavefunction methods without explicit triple excitations, such as CC2 or CCSD. \cite{loos2019reference,docasal2023classification,kossoski2024reference} 

Double excitations within the BSE@$GW$ procedure require an explicit frequency-dependent kernel. \cite{strinati1988application,rohlfing2000electron,romaniello2009double,sangalli2011double,loos2020dynamical,authier2020dynamical,bintrim2022full}
Furthermore, depending on the degree of self-consistency in the $GW$ step, \cite{vanSchilfgaarde2006quasiparticle,bruneval2006effect,blase2011first,lischner2014effects} the resulting excitation energies exhibit dependence on the chosen mean-field starting point, \cite{gui2018accuracy} albeit much reduced compared to the exchange-correlation functional dependence of (linear-response) time-dependent density-functional theory (TDDFT). \cite{rung1984,ulrich2012time,dreu2004,blase2018bethe,maitra2022double} 
Recently, BSE@$GW$ has also been extended to non-particle number conserving excitations. \cite{marie2024anomalous,marie2025anomalous}
For a more exhaustive overview of the BSE@$GW$ method, we refer to the thorough reviews and perspectives of Refs.~\onlinecite{onida2002electronic,ping2013electronic,leng2016gw,blase2018bethe,blase2020bethe}.

The success of the BSE@$GW$ method for excitation energies makes it an increasingly attractive approach for computing a broad range of excited-state properties (dipole moments, polarizabilities, etc.), \cite{himmelsbach2024excited,rauwolf2024non} adiabatic excitation energies, fluorescence energies, reorganization energies, and potentially non-adiabatic couplings for non-adiabatic molecular dynamics.  
Unfortunately, the fully analytic determination of properties, such as excited-state nuclear gradients and dipole moments, has not been achieved yet. 

While several approximate gradient formulations have been proposed in the literature, \cite{ismail2003excited,villalobos2023lagrangian} these approaches often rely on approximations to the derivatives of the quasiparticle (QP) energies and/or the effective (screened) interaction employed to construct the BSE kernel. 
The impact of such approximations in practical applications remains to be systematically assessed.
Notably, finite-difference calculations of excited-state geometries, dipole moments, and potential energy surfaces have already demonstrated the promise of the BSE@$GW$ method in accurately describing excited-state processes. \cite{ismail2003excited,kaczmarski2010diabatic,caylak2021excited,knysh2022modeling,knysh2023excited,knysh2023exploring}

Building on the recently established connection between $GW$ and CC theory, \cite{tolle2023exact} as well as the derivation of fully analytic $G_0W_0$ gradients, \cite{toelle2024fully} we present, for the first time, fully analytic nuclear gradients and molecular properties within the BSE$@G_0W_0$ framework.
Given the variety of approximations commonly employed in formulating the BSE@$GW$ working equations, we systematically examine various variants of the BSE@$G_0W_0$ method.

% theory 
\textbf{Theory:} The Bethe--Salpeter equation is defined as \cite{salpeter1951relativistic,strinati1988application,martin2016interacting}
\begin{multline} 
    L(12;1'2') = L_0(12;1'2')  
    \\ 
    + \int d(343'4') L_0(13';1'3) K(34;3'4') L(4'2;42'),
    \label{eq:BSEResponse}
\end{multline}
where $L$ denotes the interacting response function, $L_0(12;1'2') = G(12')G(21')$ its non-interacting counterpart, and the so-called BSE kernel is
\begin{equation} \label{eq:K}
    K(34';3'4) = \frac{\delta \Sigma(33')}{\delta G(44')}. 
\end{equation}
In Eq.~\eqref{eq:K}, $G$ and $\Sigma$ are the one-body Green's function and the self-energy, respectively.
Combined space-spin-time variables are denoted as integer numbers.

In the case of the $GW$ approximation for the self-energy and the static approximation for the BSE kernel, $K$ is approximated as \cite{strinati1988application,rohlfing2000electron}
\begin{equation} \label{eq:GWkernel}
\begin{split}
    \frac{\delta \Sigma(33')}{\delta G(44')} 
    \approx & - \ii \delta(33') \delta(44') v(34) \delta(t_3,t_4)
    \\
    & + \ii \delta(34)\delta(3'4')W(3^+3') \delta(t^+_3,t_{3'}),
\end{split}
\end{equation}
where $v$ is the (instantaneous) bare Coulomb operator and 
\begin{equation}
    \Sigma(33') = \ii G(33') W(3^+3')
\end{equation}
is the $GW$ self-energy with $3^+$ denoting an infinitesimal positive shift in time, i.e., $t_3^+ = t_3 + \delta$ $(\delta \to 0^+)$.
Equation \eqref{eq:GWkernel} implies that contributions arising from $\fdv*{W}{G}$ are neglected. \cite{strinati1988application,yamada2022development,quintero2022connections}
In Eq.~\eqref{eq:GWkernel}, the first term corresponds to the direct contribution, while the second term introduces exchange and correlation via the screened Coulomb potential $W$ in its static limit.
For additional details about the $GW$ approximation, the reader is referred to Refs.~\onlinecite{aryasetiawan1998gw,reining2018gw,golze2019gw,marie2023gw}.

Assuming real-valued orbitals and the approximations of Eq.~\eqref{eq:GWkernel}, the poles of the BSE [see Eq.~\eqref{eq:BSEResponse}] are commonly determined after projection in the basis of single (de)excitations through the following linear eigenvalue problem:
\begin{equation} \label{eq:BSE}
	\bH \cdot \bV_\nu = \Omega_\nu \bV_\nu,
\end{equation} 
or, more explicitly,
\begin{equation} \label{eq:BSEsta}
    \begin{pmatrix}
        \bA & \bB \\
        -\bB & -\bA
    \end{pmatrix}
    \cdot
    \begin{pmatrix}
        \bX_\nu \\
        \bY_\nu
    \end{pmatrix}
    =  
    \Omega_\nu
    \begin{pmatrix}
        \bX_\nu \\
        \bY_\nu
    \end{pmatrix},
\end{equation}
where $\nu$ labels the excited state and the elements of the (anti)resonant and coupling blocks, for singlet excited states, are given by
\begin{subequations}
\begin{align}
	\label{eq:Asta} 
    A_{ia,jb} & = \qty( \eps_a^\QP - \eps_i^\QP ) \delta_{ij} \delta_{ab} + 2 (ia|jb) - W_{ba,ji}, 
	\\
	\label{eq:Bsta}
    B_{ia,jb} & = 2 (ia|bj) - W_{ja,bi}, 
\end{align} 
\end{subequations}
while the elements of static screening read
\begin{equation} \label{eq:Wsta}
	W_{rp,qs}
	= (pr|qs) - 4 \sum_{\tI} \frac{(pr|\tI) (\tI|qs)}{\Omega_{\tI}^\RPA},
\end{equation}
with 
\begin{equation} \label{eq:excitation_density}
	(pq|\tI) = \sum_{I}  (pq|I) \qty( X_{I\tI}^\RPA + Y_{I\tI}^\RPA ).
\end{equation}
Here, the quasiparticle approximation is enforced, which means that the weight of each $GW$ quasiparticle energy, $\eps_p^\QP$, is set to $1$ \cite{Reining2016LinearResponse}
In the following, the $GW$ quasiparticle energies are denoted as $\eps_p^\QP$ and the corresponding mean-field orbital energies as $\eps_p$.
Furthermore, we employ the common notation for occupied ($i,j,k,\ldots$), virtual ($a,b,c,\ldots$), and general orbital ($p,q,r,s,\ldots$) indices. 
Capital Latin letters ($I,J,K,\ldots$) denote combined particle-hole indices, e.g., $I = ia$, while ($\tI, \tJ, \dots$) denote RPA excitation indices (\textit{vide infra}). 
The two-electron integrals $(pq|rs)$ are given in Mulliken's notation, i.e., $(11|22)$.
In this work, we also consider the Tamm-Dancoff approximation (TDA) of the static BSE presented in Eq.~\eqref{eq:BSEsta}, which is given as
\begin{equation} \label{eq:BSEtda}
    \bA \cdot \bX_\nu = \Omega_\nu \bX_\nu.
\end{equation}
The TDA consists of setting the coupling block $\bB$ to zero, resulting in a Hermitian eigenvalue problem, and it removes potential instabilities.

As readily seen in Eqs.~\eqref{eq:Asta} and \eqref{eq:Bsta}, the static BSE eigenproblem closely resembles the Casida equations of TD-DFT. \cite{casida1995time} 
The key differences are that the Kohn--Sham eigenvalues are replaced by $GW$ quasiparticle energies, the direct term remains unchanged, exact HF exchange is used instead of DFT exchange, and the correlation kernel originates from the screened Coulomb interaction $W$ rather than from a density functional approximation.

The (direct) RPA eigenvalues and eigenvectors [see Eqs.~\eqref{eq:Wsta} and \eqref{eq:excitation_density}] are computed by solving the following linear eigenvalue problem
\begin{equation} \label{eq:RPA}
    \begin{pmatrix}
        \bA^\RPA & \bB^\RPA 
        \\
        -\bB^\RPA & -\bA^\RPA 
    \end{pmatrix}
    \cdot
    \begin{pmatrix}
        \bX^\RPA_{\tI}
        \\
        \bY^\RPA_{\tI}
    \end{pmatrix}
    =  
    \Omega_{\tI}^\RPA
    \begin{pmatrix}
        \bX^\RPA_{\tI}
        \\
        \bY^\RPA_{\tI}
    \end{pmatrix}, 
\end{equation}
(with $\Omega_{\tI}^\RPA > 0$) and the elements of the RPA blocks are
\begin{subequations}
\begin{align}
	\label{eq:ARPA} 
	A_{IJ}^\RPA & \equiv A_{ia,jb}^\RPA = \qty( \eps_a - \eps_i ) \delta_{ij} \delta_{ab} + 2 (ia|jb),
	\\
	\label{eq:BRPA}
    B_{IJ}^\RPA & \equiv B_{ia,jb}^\RPA = 2 (ia|bj). 
\end{align} 
\end{subequations}
As in the BSE case, the TDA can be enforced at the RPA level by setting $\bB^\RPA = \bO$:
\begin{equation} \label{eq:TDA}
	\bA^\RPA \cdot \bX^\RPA_\tI = \Omega_\tI^\RPA \bX^\RPA_\tI.
\end{equation}
In this case, the resulting screening is referred to as TDA screening.
Conversely, when the ``full'' RPA equations are solved, the corresponding screening is termed RPA screening.

The BSE@$G_0W_0$ energy associated with the $\nu$th excited state is simply defined as
\begin{equation} \label{eq:BSE_energy}
    E_\nu = E_0 + \Omega_\nu,
\end{equation}
where $E_0$ is the $G_0W_0$ ground-state energy and $\Omega_\nu$ is the $\nu$th BSE excitation energy provided by the diagonalization of the BSE Hamiltonian \eqref{eq:BSE}.

The ground-state energy $E_0$ is given by the well-known plasmon (or trace) formula derived from the Klein functional \cite{klein1961perturbation}
\begin{equation} \label{eq:plasmon}
	E_0 = E_0^\HF + \frac{1}{2} \sum_\tI \Omega_\tI^\RPA - \frac{1}{2} \Tr(\bA^\RPA), 
\end{equation}
where $E^\text{HF}_0$ denotes the HF ground-state energy.
Consequently, if the TDA is enforced at the RPA level, we have $E_0 = E_0^\HF$, as the correlation part vanishes.

In this work, we reformulate the RPA eigenvalue problem [see Eq.~\eqref{eq:RPA}] and $G_0W_0$ energy expression [see Eq.~\eqref{eq:plasmon}], through the (direct-ring) unitary CC doubles (drUCCD) procedure, \cite{tolle2023exact,tolle2024ab,toelle2024fully} so that 
\begin{equation}
    \qty[ e^{\bt^\dagger} \cdot \qty( \bA^\RPA + \bB^\RPA ) \cdot e^{\bt} ] \bU_\tI = \Omega_\tI^\RPA \bU_\tI,
    \label{eq:RPAdrUCCD}
\end{equation}
with $\bU_\tI$ denoting the excitation vector, and $\bt$ the drUCCD amplitudes (see SI,  Sec.~I). 
This enables the evaluation of the response of RPA excitation energies and excitation vectors to an external perturbation through the inclusion of the amplitude response, an essential component in the calculation of nuclear gradients for BSE@$G_0W_0$.

Derivatives are then obtained from the following Lagrangian \cite{furc2002,villalobos2023lagrangian} 
\begin{equation} \label{eq:lagrangian}
\begin{split}
    \cL_\nu
    & = E_0 + \cE_\nu[\boldsymbol{\eps}^\QP]
    + \sum_{IJ}Z_{IJ}J_{IJ}  
    \\
    & + \sum_{p > q}\lambda_{pq} F_{pq}  
    + \sum_{pq} M_{pq} \qty( S_{pq} - \delta_{pq} ),
\end{split}
\end{equation}
where $J_{I J}$ denotes drUCCD amplitude optimization condition [see SI, Eq.~(S3)], $F_{pq}$ and $S_{pq}$ are the elements of the Fock and overlap matrix in the orbital basis, and 
\begin{equation} \label{eq:ExcLagrang}
    \cE_\nu[ \boldsymbol{\eps}^\QP ] 
    = \T{\bV_{\nu}} \cdot \bH[\boldsymbol{\eps}^\QP;\Omega_\nu] \cdot \bV_{\nu} 
    - \Omega_\nu \qty( \T{\bV_{\nu}} \cdot \bV_{\nu} - 1 )
\end{equation} 
denotes the BSE excited-state functional, \cite{furc2002} which itself depends on the quasiparticle energies $\boldsymbol{\eps}^\QP = \{ \eps_p^\QP \}$ [see Eq.~\eqref{eq:Asta}].

The Lagrange multipliers $Z_{I J}$ enforce the drUCCD amplitude equation [see SI, Eq.~(S3)], \cite{tolle2023exact,tolle2024ab,toelle2024fully} while the Lagrange multipliers $\lambda_{pq}$ and $M_{pq}$ take care of orbital stationarity and orthonormality, respectively.
Note that, in contrast to the previous work on analytic $G_0W_0$ gradients, \cite{toelle2024fully} we also employ the diagonal approximation for the $GW$ quasiparticle energies. \cite{bintrim2021full} 
As a result, because the quasiparticle energies are not invariant under orbital rotation, the occupied-occupied and virtual-virtual blocks of $\lambda_{pq}$ have to be explicitly considered. 
This is in contrast to Ref.~\onlinecite{villalobos2023lagrangian}.

When the TDA is applied at the RPA level for computing the screened Coulomb interaction, i.e., for both the $GW$ quasiparticle energies and the BSE kernel, the Lagrangian \eqref{eq:lagrangian} simplifies significantly to:
\begin{equation} \label{eq:lagrangian_TDA}
    \begin{split}
    	\cL_\nu 
    	& = E^\text{HF}_0 + \cE_\nu[\boldsymbol{\eps}^\QP]
		\\
        & + \sum_{p > q}\lambda_{pq} F_{pq}  
        + \sum_{pq} M_{pq} \qty( S_{pq} - \delta_{pq} ),
	\end{split}
\end{equation}

Once $\cL_\nu$ is stationary with respect to all its parameters, the nuclear gradient for the $\nu$th excited state is calculated as 
\begin{equation} \label{eq:GradLagrang}
    \bG_\nu = \pdv{\cL_\nu}{\bR},
\end{equation}
where $\bR$ denotes the set of nuclear coordinates.

In this work, we investigate four variants of the BSE@$G_0W_0$ scheme, which differ in how the $G_0W_0$ screening is computed and how the BSE eigenvalue problem is solved.
These variants are listed in Table \ref{tab:GWBSEvariants}, along with their respective abbreviations and overall computational scaling, which is either $\order*{N^6}$ or $\order*{N^7}$ in the current implementation ($N$ being the number of orbitals).  
The working equations used to evaluate Eq.~\eqref{eq:GradLagrang} for each variant are provided in the Supporting Information (SI) Sec.~II.

% table I 
\begin{table*}
    \caption{Different BSE@$G_0W_0$ variants considered in this work, together with their corresponding abbreviation and computational scaling.}
    \begin{ruledtabular}
    \begin{tabular}{cccccc}
        & \mc{2}{c}{BSE@$GW$ variant} \\
        \cline{2-3}
        \# 	& Screening in $G_0W_0$ \& BSE kernel & BSE problem & Abbreviation & Lagrangian & Scaling \\
        \hline
        1	& RPA [Eq.~\eqref{eq:RPA}] & full BSE [Eq.~\eqref{eq:BSEsta}] 
        	& BSE@$GW$ & Eq.~\eqref{eq:lagrangian} & $\order{N^7}$ [SI, Eq.~(S41j) for $\boldsymbol{\eps}^\QP$]
		\\
        2 	& RPA [Eq.~\eqref{eq:RPA}] & TDA BSE [Eq.~\eqref{eq:BSEtda}] & BSE$_\TDA$@$GW$ 
        	& Eq.~\eqref{eq:lagrangian} & $\order{N^7}$ [SI, Eq.~(S41j) for $\boldsymbol{\eps}^\QP$]
		\\
        3 	& TDA [Eq.~\eqref{eq:TDA}] & full BSE [Eq.~\eqref{eq:BSEsta}] & BSE@$GW_\TDA$ 
        	& Eq.~\eqref{eq:lagrangian_TDA} & $\order{N^6}$ [SI, Eq.~(29l)]
        \\
        4 	& TDA [Eq.~\eqref{eq:TDA}] & TDA BSE [Eq.~\eqref{eq:BSEtda}] & BSE$_\TDA$@$GW_\TDA$ 
        	& Eq.~\eqref{eq:lagrangian_TDA} & $\order{N^6}$ [SI, Eq.~(29l)]
        \\
    \end{tabular}
    \end{ruledtabular}
    \label{tab:GWBSEvariants}
\end{table*}

% computational details
\textbf{Computational details:} A first pilot implementation of the various BSE@$GW$ nuclear gradient expressions has been developed in a Python code that heavily relies on functionalities provided by \textsc{PySCF}. \cite{sun2018pyscf,sun2020recent}
Geometry optimizations are performed via the \textsc{geomeTRIC} \cite{wang2016geometry} interface to \textsc{PySCF}, using the following modified convergence thresholds: the maximum and root-mean-square components of the gradient (\verb|convergence_gmax| and \verb|convergence_grms|) are set to \SI{d-4}{\hartree\per\bohr}, while the maximum and root-mean-square displacements in atomic coordinates (\verb|convergence_dmax| and \verb|convergence_drms|) are set to \SI{d-4}{\angstrom}. 
The energy convergence criterion (\verb|convergence_energy|) is set to \SI{d-6}{\hartree}.

\begin{figure}[b!]
    \includegraphics[width=\linewidth]{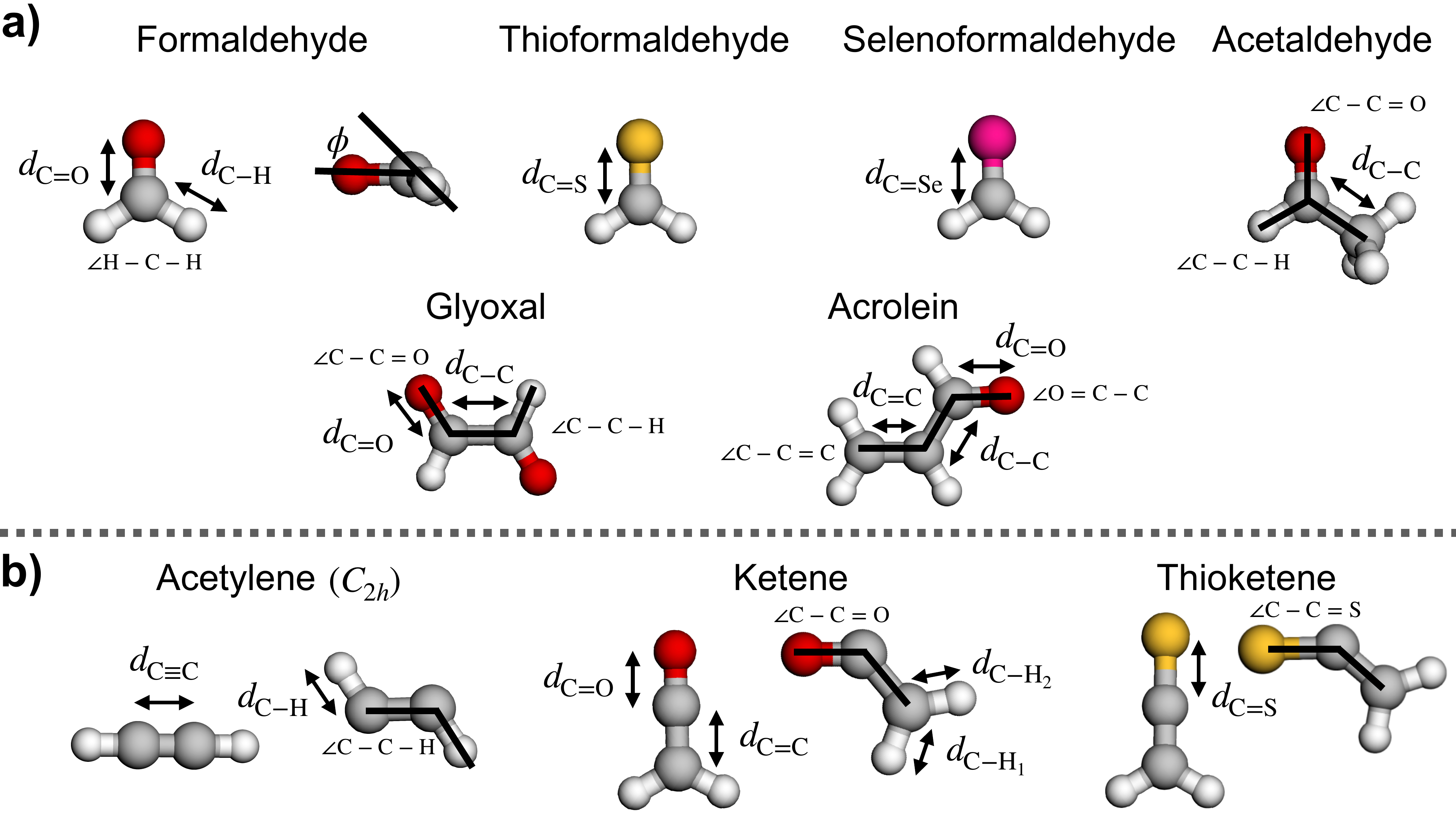}
    \caption{Molecular structures and geometry parameters considered in this work [gray: C, red: O, white: H, yellow: S, purple: Se]: a) $n\to\pi^*$ transitions; b) $\pi\to\pi^*$ transitions.}
    \label{fig:StrucIllustration}
\end{figure}

Below, we compare geometrical parameters for excited-state structures obtained from the different BSE@$GW$ variants (see Table \ref{tab:GWBSEvariants}) and wavefunction-based approaches.
The molecular structures and the corresponding geometrical parameters considered in the following are illustrated in Fig.~\ref{fig:StrucIllustration}, with reference geometries taken from Ref.~\onlinecite{budzak2017accurate}.
Excited-state BSE geometry optimizations were performed starting from both the ground- and excited-state geometries of Ref.~\onlinecite{budzak2017accurate}. 
Ground-state BSE optimizations, based on the HF or RPA energy functionals [see Eq.~\eqref{eq:plasmon}], consistently use the ground-state geometry as the initial structure.
The geometrical changes discussed in for the various BSE@$GW$ approaches are defined relative to the ground-state BSE geometry optimized based on the RPA energy functional for BSE@$GW$ and BSE$_\TDA$@$GW$, or relative to the HF optimized geometry for BSE@$GW_\text{TDA}$ and BSE$_\text{TDA}$@$GW_\text{TDA}$.

Only singlet excited states are considered in this work, although an extension to the triplet manifold is straightforward.
All reported geometries correspond to (local) minima on the excited-state potential energy surface, as verified by numerical frequency analysis of the excited-state Hessian.
For acetaldehyde, two distinct local minima are found for both BSE@$GW$ and BSE$_\TDA$@$GW$, depending on whether the optimization starts from the reference ground- or excited-state structure. 
The energy difference between these two conformers is about \SI{0.03}{\eV} at the BSE level [see Eq.~\eqref{eq:BSE_energy}].  
We therefore restrict our analysis to the lower-energy conformer. 
For acetylene, we enforce the excited-state geometry to be of $C_{2\text{h}}$ symmetry, which corresponds to a saddle point for both BSE@$GW$ and BSE$_\TDA$@$GW$.

In all calculations, the broadening parameter $\eta$ is set to zero.
The $G_0W_0$ step is performed by solving the full quasiparticle equation (i.e., without linearization), making use of the frequency-independent $G_0W_0$ formulation of Ref.~\citenum{bintrim2021full}.
We restrict our study to closed-shell systems and consistently employ a restricted formalism.
Although it is common practice to adopt Kohn--Sham orbitals and energies as the starting point, in this work, we systematically use HF orbitals and their corresponding eigenvalues.
Generalization to Kohn--Sham starting points is left for future investigations.

In BSE@$GW$ calculations, quasiparticle corrections are applied to all occupied HF orbital energies except for the core orbitals. The number of core orbitals for the different elements is obtained from the \texttt{chemcore} function of \textsc{PySCF}.
We also include the $N_\text{occ} + 10$ (number of occupied orbitals plus ten) lowest unoccupied quasiparticle states.
Furthermore, we verify that each included quasiparticle state has a quasiparticle weight greater than 0.5; otherwise, the corresponding state is discarded, and the HF orbital energy is used in its place. \cite{Reining2016LinearResponse}

All gradient calculations employ the def2-TZVPP basis set, \cite{weigend2005balanced} whereas the \textit{aug}-cc-pVTZ basis set \cite{dunning1989gaussian,kendall1992electron} is used for the determination of excitation energies.
Throughout the self-consistent field (SCF) procedure is converged to an energy change below \SI{d-10}{\hartree}, while drUCCD amplitudes are iterated until the norm of the amplitude update satisfies $\norm{\Delta \bt} < 10^{-8}$.
The Davidson solver used for computing both $G_0W_0$ and BSE eigenvalues is converged to \SI{d-8}{\hartree}, and the LGMRES\cite{baker2005lgmres} procedure used for determining the $Z$-Lagrange multipliers (SI, Sec.~II H) is converged below \num{e-5} (SI, Sec.~II H).

The validity of our fully analytic BSE@$GW$ gradient implementation has been verified by comparison with numerical (finite-difference) gradients for a set of five diatomic molecules at their equilibrium bond length. 
As detailed in Supporting Information (SI) Sec.~III, we observe excellent agreement between analytic and numerical results across all considered BSE@$GW$ variants, confirming the correctness of our derivations and implementation.

% excited-state geometries
\textbf{Excited-state geometries:}
% figure 1
% interesting molecules
As a first set of molecular test systems for assessing the accuracy of BSE@$GW$ excited-state gradients, we consider formaldehyde, thioformaldehyde, selenoformaldehyde, acetaldehyde, acrolein, and glyoxal (see upper panel of Fig.~\ref{fig:StrucIllustration}). 
For these systems, highly accurate reference geometries from CC methods are available: CC3 for formaldehyde, thioformaldehyde, and selenoformaldehyde, and CCSDR(3) for acetaldehyde, acrolein, and glyoxal. \cite{budzak2017accurate}
Our analysis focuses on the lowest singlet excited state of $n \to \pi^*$ character.

To evaluate the quality of the predicted excited-state geometries of formaldehyde, thioformaldehyde and selenoformaldehyde, we examine three key structural parameters: the \ce{C=M} bond length (where $\text{M} = \text{O}$, S, or Se), the \ce{C-H} bond length, and the \ce{H-C-H} bond angle (see Fig.~\ref{fig:StrucIllustration}).
For formaldehyde, we also consider the out-of-plane angle $\phi$, which significantly differs from zero in the excited state.
The computed geometrical parameters, along with results from second-order wavefunction-based methods [ADC(2), CC2, and CCSD], are compiled in Table \ref{tab:Formaldehyde}. 
Across all systems, the $S_0 \to S_1$ excitation results in an elongation of the \ce{C=M} double bond and a slight contraction of the \ce{C-H} bond. 
In formaldehyde, the excitation also induces a significant out-of-plane distortion, reflected by a nonzero dihedral angle. The resulting out-of-plane angle predicted by CC3 is \SI{37.4}{\degree}.

For all compounds, the predicted changes in the \ce{C=M} double bond length show good agreement with the CC3 reference, especially for the BSE@$GW$ and BSE$_\TDA$@$GW$ variants. 
In other words, the TDA at the BSE level has little impact on the structural parameters, as expected for the present molecular systems, which do not exhibit plasmon-like excitations.
This trend is very systematic over all systems considered in the study.
Notably, the bond length changes predicted by RPA-based BSE are systematically smaller relative to the $GW_\TDA$-based BSE variants. 

When the screening is computed within the TDA [$GW_\TDA$ variants], the out-of-plane angle $\phi$ is systematically zero. 
For BSE@$GW$ and BSE$_\TDA$@$GW$, the out-of-plane angle is smaller than the reference CC3 value and comparable to the angle obtained with ADC(2).
The vanishing out-of-plane angle in the $GW_\TDA$ variants may be attributed to the HF reference energy in the total energy functional [see Eq.~\eqref{eq:plasmon}].
Precisely disentangling the origin of the differences in the observed structural changes is, however, difficult and will be the focus of future work. 

% table 3
\begin{table*}[h!]
    \caption{Changes in geometrical parameters upon excitation [$\Delta$\ce{C=M} (\ce{C=M} bond length, in \si{\AA}), $\Delta$\ce{C-H} (\ce{C-H} bond length, in \si{\AA}), $\Delta\angle$ (\ce{H-C-H} bond angle, in degrees), and $\Delta\phi$ (\ce{H2C=M} out-of-plane angle, in degrees); see Fig.~\ref{fig:StrucIllustration}] for the lowest singlet excited state  [$^1 A_2$ ($n \to \pi^*$)] of formaldehyde ($\text{M} = \text{O}$), thioformaldehyde ($\text{M} = \text{S}$), and selenoformaldehyde ($\text{M} = \text{Se}$) computed in the def2-TZVPP basis set.}
    \label{tab:Formaldehyde}
    \begin{ruledtabular}
    \begin{tabular}{l c c c c c c c c c c}
        & \multicolumn{4}{c}{Formaldehyde} & \multicolumn{3}{c}{Thioformaldehyde} & \multicolumn{3}{c}{Selenoformaldehyde} \\
        \cline{2-5} \cline{6-8} \cline{9-11}
        Method & $\Delta$\ce{C=O} & $\Delta$\ce{C-H} & $\Delta\angle$ & $\Delta$$\phi$ & $\Delta$\ce{C=S} & $\Delta$\ce{C-H} & $\Delta\angle$ & $\Delta$\ce{C=Se} & $\Delta$\ce{C-H} & $\Delta\angle$\\
        \hline
        BSE@$GW$ & 0.126&$-0.023$& 5.0&22.9& 0.088&$-0.005$&3.0&0.095&$-0.004$&4.0	\\
        BSE$_\TDA$@$GW$ & 0.127&$-0.023$& 6.0&18.4& 0.089&$-0.005$&3.0&0.095&$-0.004$&4.0\\
        BSE@$GW_\TDA$ & 0.132&$-0.027$& 7.0&0.1 & 0.108&$-0.005$&3.0&0.131&$-0.003$&3.0	\\
        BSE$_\TDA$@$GW_\TDA$ & 0.131&$-0.027$& 7.0&0.0 & 0.107&$-0.004$&3.0&0.130&$-0.003$&3.0\\
        \hline
        ADC(2)\fnm[1] & 0.171 &$-0.016$& 7.0& 19.8& 0.113& $-0.004$& 4.8 & 0.132& $-0.002$& 4.7\\
        CC2\fnm[1]    & 0.138 &$-0.013$& 4.6& 30.8& 0.091& $-0.004$& 4.6 & 0.098& $-0.002$& 4.5\\
        CCSD\fnm[1]   & 0.104 &$-0.011$& 2.6& 31.8& 0.074& $-0.004$& 3.5 & 0.077& $-0.003$& 3.3\\
        CC3\fnm[1]    & 0.122 &$-0.011$& 2.0& 37.4& 0.093& $-0.005$& 4.2 & 0.095& $-0.003$& 3.8\\ 
    \end{tabular}
    \fnt[1]{Reference values taken from Ref.~\onlinecite{budzak2017accurate}.}
    \end{ruledtabular}
\end{table*}

The results for acetaldehyde, acrolein, and glyoxal are presented in Tables \ref{tab:Acetaldehyde}, \ref{tab:Acrolein}, and \ref{tab:Glyoxal}, respectively.
In the case of acetaldehyde, all wavefunction-based methods [CCSD and CCSDR(3)] predict a non-zero out-of-plane angle $\phi$ in the lowest excited state.
In contrast, all BSE@$GW$ variants yield a strictly planar excited-state geometry, with $\phi = \SI{0}{\degree}$. 
Additionally, BSE@$GW$ predicts a shortening of the \ce{C-C} bond upon excitation, in disagreement with the elongation observed in the CCSD and CCSDR(3) reference geometries. 
Apart from these discrepancies, the remaining structural changes are qualitatively reproduced by the BSE@$GW$ approaches.
The BSE@$GW$ variants generally reproduce the structural changes in acrolein and glyoxal qualitatively well, with the exception of the \ce{C-C=C} angle in acrolein. 
Unlike the CCSDR(3) reference, both BSE@$GW_\TDA$ and BSE$_\TDA$@$GW_\TDA$ incorrectly predict a decrease in this angle. 
Notably, BSE$_\TDA$@$GW$ and BSE$_\TDA$@$GW$ match or outperform CCSD in predicting the elongation of the \ce{C=O} bond and the contraction of the \ce{C-C} bond in both acrolein and glyoxal, when compared against the CCSDR(3) benchmark.

% table 4
\begin{table*}
    \caption{Changes in geometrical parameters upon excitation [$\Delta$\ce{C=O} and $\Delta$\ce{C-C} (bond lengths, in \si{\AA}), $\Delta$\ce{C-C=O} and $\Delta$\ce{C-C-H} (bond angles, in degrees), and $\Delta\phi$ (out-of-plane angle, in degrees); see Fig.~\ref{fig:StrucIllustration}] for the lowest singlet excited state [$^1 A''$ ($n \to \pi^*$)] of acetaldehyde computed in the def2-TZVPP basis set.}
    \label{tab:Acetaldehyde}
    \begin{ruledtabular}
    \begin{tabular}{l c c c c c}
        Method & $\Delta$\ce{C=0} & $\Delta$\ce{C-C} & $\Delta$\ce{C-C=O} & 
        $\Delta$\ce{C-C-H} & $\Delta\phi$ \\
        \hline
        BSE@$GW$  & 0.142&$-0.023$&$-6.5$&10.2&0.0 \\
        BSE$_\TDA$@$GW$  & 0.142&$-0.023$&$-6.6$&10.2&0.0 \\
        BSE@$GW_\TDA$  & 0.157&$-0.035$&$-7.0$&10.6&0.0\\
        BSE$_\TDA$@$GW_\TDA$  & 0.156&$-0.035$&$-7.0$&10.6&0.0\\
        \hline
        CCSD\fnm[1]        & 0.107 & 0.006 & $-8.2$ & 4.2 & 34.8 \\
        CCSDR(3)\fnm[1]    & 0.122 & 0.005 & $-9.1$ & 3.9 & 38.7\\ 
    \end{tabular}
	\fnt[1]{Reference values taken from Ref.~\onlinecite{budzak2017accurate}.}
    \end{ruledtabular}
\end{table*}

\begin{table*}
    \caption{Changes in geometrical parameters upon excitation [$\Delta$\ce{C=O}, $\Delta$\ce{C-C} and $\Delta$\ce{C=C} (bond lengths, in \si{\AA}), $\Delta$\ce{O=C-C} and $\Delta$\ce{C-C=C} (bond angles, in degrees); see Fig.~\ref{fig:StrucIllustration}] for the lowest singlet excited state [$^1 A''$ ($n \to \pi^*$)] of acrolein computed in the def2-TZVPP basis set.}
    \label{tab:Acrolein}
    \begin{ruledtabular}
    \begin{tabular}{l c c c c c}
        Method & $\Delta$\ce{C=0} & $\Delta$\ce{C-C} & $\Delta$\ce{C=C}&$ \Delta$\ce{O=C-C} & 
        $\Delta$\ce{C-C=C} \\
        \hline
        BSE@$GW$  & 0.123 & $-$0.087 & 0.039 & 0.5 & 1.9 \\
        BSE$_\TDA$@$GW$  &  0.123 & $-$0.087 & 0.040 & 0.7 & 1.9 \\
        BSE@$GW_\TDA$  & 0.125 & $-$0.108 & 0.057 & 2.8 & $-$0.3 \\
        BSE$_\TDA$@$GW_\TDA$  & 0.124 & $-$0.109 & 0.058 & 3.0 & $-$0.3\\
        \hline
        CCSD\fnm[1]  &  0.099 & $-$0.063 & 0.025 & 0.4 & 3.2 \\
        CCSDR(3)\fnm[1] & 0.111 & $-$0.081 & 0.037 & 0.9 & 3.0\\ 
    \end{tabular}
	\fnt[1]{Reference values taken from Ref.~\onlinecite{budzak2017accurate}.}
    \end{ruledtabular}
\end{table*}

\begin{table*}
    \caption{Changes in geometrical parameters upon excitation [$\Delta$\ce{C=O}, $\Delta$\ce{C-C} and $\Delta$\ce{C-H} (bond lengths, in \si{\AA}), $\Delta$\ce{C-C=O} and $\Delta$\ce{C-C-H} (bond angles, in degrees); see Fig.~\ref{fig:StrucIllustration}] for the lowest singlet excited state [$^1 A_u$ ($n \to \pi^*$)] of glyoxal computed in the def2-TZVPP basis set.}
    \label{tab:Glyoxal}
    \begin{ruledtabular}
    \begin{tabular}{l c c c c c}
        Method & $\Delta$\ce{C=0} & $\Delta$\ce{C-C} & $\Delta$\ce{C-H}&$ \Delta$\ce{C-C=O} & 
        $\Delta$\ce{C-C-H} \\
        \hline
        BSE@$GW$  & 0.030 & $-$0.033 & $-$0.011 & 2.4 & $-$1.3 \\
        BSE$_\TDA$@$GW$  & 0.029 & $-$0.034 & $-$0.010 & 2.5& $-$1.3  \\
        BSE@$GW_\TDA$  & 0.028 & $-$0.031 & $-$0.014 & 1.8 & $-$1.0 \\
        BSE$_\TDA$@$GW_\TDA$  & 0.028 & $-$0.031 & $-$0.013 & 1.9 & $-$1.0 \\
        \hline
        CCSD\fnm[1]  &  0.025 & $-$0.027 & $-$0.006 & 2.3 & $-$1.3 \\
        CCSDR(3)\fnm[1] & 0.030 & $-$0.035 & $-$0.006 & 2.1 & $-$0.8 \\ 
    \end{tabular}
	\fnt[1]{Reference values taken from Ref.~\onlinecite{budzak2017accurate}.}
    \end{ruledtabular}
\end{table*}

Next, we explore the changes in geometrical parameters upon excitation for acetylene (which retains $C_{2\text{h}}$ symmetry in the excited state), ketene, and thioketene (see lower panel of Fig.~\ref{fig:StrucIllustration}).
In all cases, we consider the lowest singlet excited state, which corresponds to a $\pi \to \pi^*$ transition.
The reference geometrical parameters, displayed in Fig.~\ref{fig:StrucIllustration}, are again taken from Ref.~\onlinecite{budzak2017accurate}.
The results for the various BSE@$GW$ variants and wavefunction-based methods are summarized in Table \ref{tab:Acetylene}.

In acetylene, the \ce{C#C} triple bond and \ce{C-H} single bond lengths increase in the excited state.
Furthermore, the linear structure (with $D_{\infty \text{h}}$ symmetry) is distorted upon excitation, reducing the \ce{C#C-H} angle to less than \SI{180}{\degree} and lowering the symmetry to $C_{2\text{h}}$. 

All methods considered capture the main features of the excited-state structural changes in acetylene.
Among the BSE@$GW$ variants, BSE@$GW$ and BSE$_\TDA$@$GW$ show the best overall agreement with the CC3 reference.
Interestingly, BSE@$GW$ and BSE$_\TDA$@$GW$ yield smaller deviations from CC3 than CCSD, which shows larger discrepancies than CC2 and ADC(2), likely due to favorable error cancellation in the latter two methods.

For both ketene and thioketene, the \ce{C=O} and \ce{C=S} double bonds elongate upon excitation.
Additionally, the \ce{C=C=O} and \ce{C=C=S} bond angles deviate from linearity, leading to inequivalent positions for the two hydrogen atoms (\ce{H_1} and \ce{H_2}).
Despite this asymmetry, the differences in the changes of the \ce{C-H_1} and \ce{C-H_2} bond lengths remain small (\SI{0.008}{\AA} and \SI{0.000}{\AA} for ketene, and \SI{0.007}{\AA} and \SI{0.001}{\AA} for thioketene).
In the case of ketene, BSE@$GW$ correctly predicts the correct direction of change (elongation or shortening) for both \ce{C-H} bonds.
For thioketene, all BSE@$GW$ variants qualitatively reproduce the correct trend in the bond-length changes.
For both molecules, the variants employing RPA screening generally yield smaller deviations across the various geometrical parameters. 
Again, disentangling the precise origins of these differences, whether arising from the choice of ground-state energy functional, the screening in the quasiparticles, or the BSE kernel, lies beyond the scope of this study. 

\begin{squeezetable}
\begin{table*}
    \caption[]{Changes in geometrical parameters upon excitation (see Fig.~\ref{fig:StrucIllustration}) for the lowest singlet excited state ($\pi \to \pi^*$) of acetylene [$\Delta$\ce{C#C}, $\Delta$\ce{C-H} (bond lengths, in \si{\AA}), $\Delta$\ce{C#C-H} (bond angle, in degrees)], ketene [$\Delta$\ce{C=O}, $\Delta$\ce{C=C}, $\Delta$\ce{C-H_1}, $\Delta$\ce{C-H_2} (bond lengths, in \si{\AA}), $\Delta$\ce{C=C=O} (bond angle, in degrees)], and thioketene [$\Delta$\ce{C=S}, $\Delta$\ce{C=C}, $\Delta$\ce{C-H_1}, $\Delta$\ce{C-H_2} (bond lengths, in \si{\AA}), $\Delta$\ce{C=C=S} (bond angle, in degrees)] computed in the def2-TZVPP basis set.}
    \label{tab:Acetylene}
    \begin{ruledtabular}
    \begin{tabular}{l c c c c c c c c c c c c c}
        & \multicolumn{3}{c}{Acetylene} & \multicolumn{5}{c}{Ketene} & \multicolumn{5}{c}{Thioketene} \\
        \cline{2-4} \cline{5-9} \cline{10-14}
        Method & $\Delta$C$\equiv$C & $\Delta$C$-$H & $\Delta$C$\equiv$C$-$H &
        $\Delta$\ce{C=O} & $\Delta$\ce{C=C} & $\Delta$\ce{C-H_1} & $\Delta$\ce{C-H_2} & $\Delta$\ce{C=C=O} &
        $\Delta$\ce{C=S} & $\Delta$\ce{C=C} & $\Delta$\ce{C-H_1} & $\Delta$\ce{C-H_2} & $\Delta$\ce{C=C=S} \\
        \hline
        BSE@$GW$ & 0.177&0.031&$-59.0$ &0.026&0.137&0.007&0.000&$-51.5$&0.056&0.045&0.007&0.002&$-40.9$\\
        BSE$_\TDA$@$GW$ & 0.175&0.031&$-58.5$&0.026&0.129&0.007&$-0.001$&$-49.9$&0.054&0.044&0.007&0.002&$-39.4$\\
        BSE@$GW_\TDA$ & 0.149&0.029&$-53.5$&0.017&0.142&0.006&	$-0.002$&$-46.2$&0.067&0.030&0.007&0.003&$-31.8$\\
        BSE$_\TDA$@$GW_\TDA$ & 0.142&0.028&$-52.2$&0.017&0.135&0.006&	$-0.002$&$-45.2$&0.065&0.030&0.007&0.003&$-30.5$\\
        \hline
        ADC(2)\fnm[1] & 0.164 &0.031 &$-58.0$&  0.036& 0.115 &0.008& 0.000& $-49.7$ & 0.064& 0.049 &0.007 &0.001 &$-40.8$ \\
        CC2\fnm[1]    & 0.167 &0.030 &$-58.1$& 0.036& 0.119 &0.008& 0.000& $-50.7$ &  0.052& 0.056& 0.008 &0.001 &$-42.2$\\
        CCSD\fnm[1]   & 0.152 &0.030 &$-55.9$& 0.033& 0.104& 0.009& 0.001& $-48.6$ & 0.048& 0.049& 0.008 &0.002 &$-39.9$\\
        CC3\fnm[1]    & 0.168 &0.033 &$-57.9$&0.037 &0.116& 0.008& 0.000 &$-50.5$ & 0.054& 0.058& 0.007 &0.001 &$-43.0$\\ 
    \end{tabular}
	\fnt[1]{Reference values taken from Ref.~\onlinecite{budzak2017accurate}.}
    \end{ruledtabular}
\end{table*}
\end{squeezetable}

% excited-state energies
\textbf{Excited-state energies:} 
Following the analysis of excited-state geometrical changes, we compare absorption, fluorescence, and adiabatic transition energies computed with the various BSE@$GW$ variants, based on the optimized structures presented.
Reference CC3 transition energies are taken from Ref.~\citenum{loos2019chemically} [CCSDR(3)/def2-TZVPP geometries]. 
Furthermore, we compute BSE@$GW$ vertical transition energies using the BHLYP [$50\%$ of B88 exchange, \cite{beck1988} $50\%$ of HF exchange, and $100\%$ of LYP correlation \cite{lee1988development}] exchange--correlation functional [BSE@$GW$@BHLYP] as an alternative mean-field starting point for BSE@$GW$ single-point calculations.
The ground-state energy functional $E_0$ in this case is Eq.~\eqref{eq:plasmon} for BSE@$GW$ and BSE$_\TDA$@$GW$, whereas BSE@$GW_\TDA$ and BSE$_\TDA$@$GW_\TDA$ use $E_0 = E_0^\text{BHLYP}$.

\begin{figure}
    \includegraphics[width=0.5\textwidth]{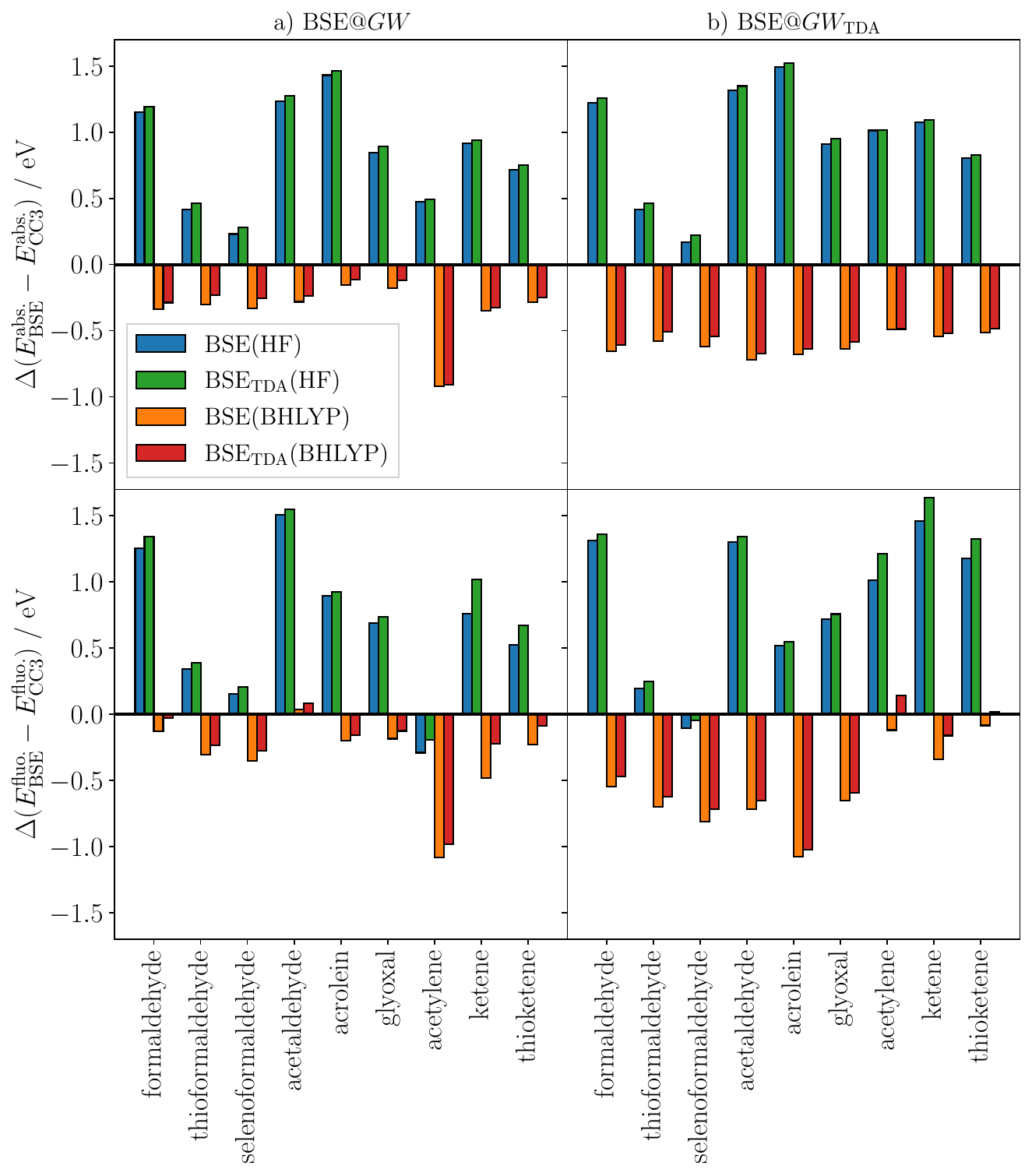}
    \caption{Deviation in absorption (abs.) and fluorescence (fluo.) transition energies ($S_0 \to S_1$), calculated using BSE@$GW$ with and without the TDA and different mean-field starting points (HF and BHLYP), relative to CC3 reference excitation energies of Ref.~\onlinecite{loos2019chemically}. 
    All excitation energies are computed in the \textit{aug}-cc-pVTZ basis set. 
    BSE@$GW$ calculations rely on the optimized geometries of this work.}
    \label{fig:BSE_Abs_Fluo_Energies}
\end{figure}

The deviations in the absorption and fluorescence transition energies are displayed in Fig.~\ref{fig:BSE_Abs_Fluo_Energies}.
Additional comparison between BSE@$GW$@BHLYP and TDDFT@BHLYP transition energies is provided in the SI Sec.~IV.
All HF-based variants mostly overestimate, whereas the BHLYP-based BSE@$GW$ variants underestimate transition energies relative to the CC3 reference. 
For both mean-field starting points, only a small difference between BSE and BSE$_\TDA$ is observed, similar to the influence of the TDA on geometrical parameters.

\begin{figure}
    \includegraphics[width=0.5\textwidth]{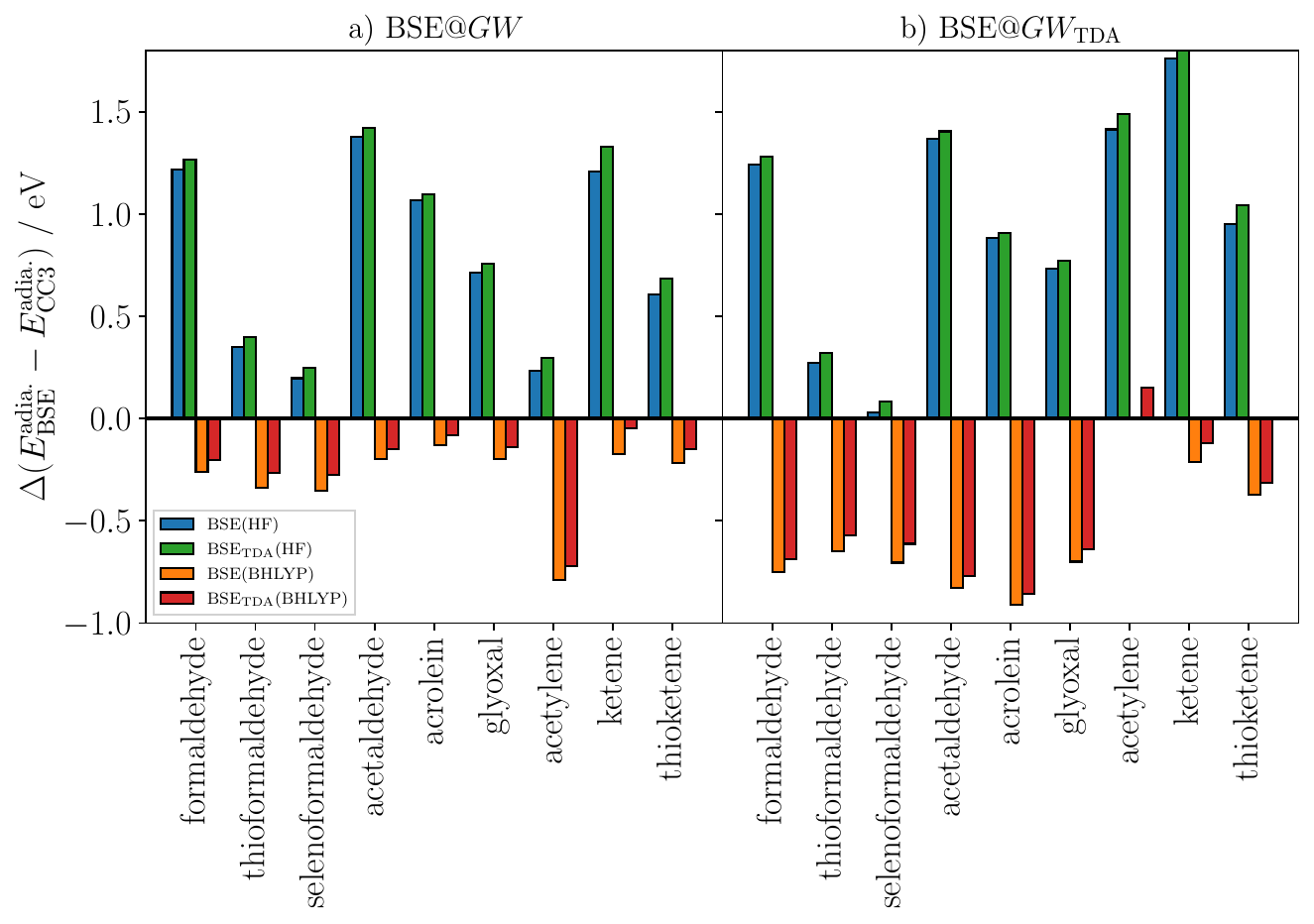}
    \caption{Deviation in adiabatic transition energies ($S_0 \to S_1$), calculated using BSE@$GW$ with and without the TDA and different mean-field starting points (HF and BHLYP), relative to CC3 reference excitation energies of Refs.~\onlinecite{loos2019chemically} and \onlinecite{budzak2017accurate} (ketene and thioketene). 
    All excitation energies are computed in the \textit{aug}-cc-pVTZ basis set. 
    BSE@$GW$ calculations rely on the optimized geometries of this work.}
    \label{fig:BSE_Abdia_Energies}
\end{figure}

The mean absolute errors (MAEs) for the absorption and fluorescence transition energies are $0.82(0.86)/0.71(0.78)$ eV for BSE@$GW$@HF and $0.35(0.30)/0.33(0.24)$ eV for BSE@$GW$@BHLYP, with the value in parentheses denoting the MAEs for BSE$_\TDA$.
Applying the TDA to the screened Coulomb interaction increases the deviations, yielding MAEs of $0.94(0.97)/0.87(0.94)$ eV for BSE@$GW_\TDA$@HF and $0.61(0.56)/0.56(0.49)$ eV for BSE@$GW_\TDA$@BHLYP.
These trends are also reflected in the adiabatic transition energies, shown in Fig.~\ref{fig:BSE_Abdia_Energies}. Specifically, the BSE@$GW$@HF/BSE@$GW_\TDA$@HF and BSE@$GW$@BHLYP/BSE@$GW_\TDA$@BHLYP approaches yield MAEs of $0.77(0.83)/0.96(1.01)$ eV and $0.30(0.23)/0.57(0.52)$ eV, respectively.

Given the good performance of BSE@$GW$@HF and BSE$_\text{TDA}$@$GW$@HF in capturing structural changes in the excited state (see above), the relatively large deviations observed in their predicted transition energies are somewhat unexpected. 
Switching the mean-field starting point to BHLYP for the transition energy calculations leads to a significant improvement in accuracy.  
To further explore this aspect, we computed absorption/fluorescence/adiabatic transition energies for acetylene using CC3 (\textit{aug}-cc-pVTZ), based on the RPA@HF/BSE@$GW$@HF optimized structures.
For this system, BSE@$GW$@BHLYP (with and without TDA) has its largest deviation.
The resulting CC3 deviations are $0.16$/$-0.03$/$-0.08$ eV, highlighting again that BSE@$GW$@HF reliably captures excited-state geometrical changes.
Accurate transition energies, however, require choosing a different mean-field starting point for BSE@$GW$.

Alternatively, it has been shown in Ref.~\onlinecite{loos2020dynamical} that the inclusion of dynamical effects in the BSE results in a red-shift in BSE@$GW$@HF excitation energies for small molecular systems, improving the overall accuracy for absorption excitation energies.
Whether similar dynamical effects also reduce the error in fluorescence and adiabatic transition energies remains an open question and will be explored in future work.
This also holds for partial or fully self-consistent $GW$ procedures. 

In summary, we derived, implemented, and validated analytic nuclear gradients for different BSE@$GW$ variants (Table \ref{tab:GWBSEvariants}), starting from a HF mean-field reference.
Subsequently, the capabilities of these different methods for describing bond length and bond angle changes in the lowest singlet excited state ($n\to\pi^*$ and $\pi\to\pi^*$) for nine prototypical small molecular systems were investigated. 

The different BSE@$GW$ variants generally capture the qualitative trends in the excited-state geometrical parameters, with  BSE@$GW$, based on full RPA screening, yields geometries in qualitative agreement with reference wavefunction-based methods [CC3/CCSDR(3)].
Notably, the $GW$-based variants with TDA screening fail to reproduce the out-of-plane angles in the excited-state geometry of formaldehyde and acetaldehyde.
The reason behind this will be investigated in a more thorough future investigation of the BSE@$GW$ excited-state properties.

In contrast to the geometrical changes, BSE@$GW$ transition energies (absorption, fluorescence, adiabatic), starting from a HF mean-field reference, tend to overestimate the CC3 reference excitation energies. 
The discrepancy is reduced when using a BHLYP mean-field reference for computing the vertical transition energies, resulting in MAEs of $0.35/0.33$ eV for the absorption and fluorescence transition energies and \SI{0.30}{\eV} for the adiabatic transition energies.
Due to the limited number of systems studied, no definitive conclusions can yet be drawn regarding the overall performance of BSE@$GW$ for excited-state properties.
In this regard, an extension of the present formalism to self-consistent $GW$ schemes, e.g., eigenvalue self-consistent $GW$, is possible and planned in the future.

To this end, we aim to reduce the computational cost associated with the BSE@$GW$ calculations [currently $\mathcal{O}(N^6)$ or $\mathcal{O}(N^7)$] by employing efficient screening techniques for the quasiparticle contributions within the BSE, as well as density-fitting, separable resolution of the identity methods and the auxiliary boson expansion. \cite{ren2012resolution,krause2017implementation,duchemin2019separable,duchemin2021cubic,liu2020all,zhou2024all,tolle2024ab}
These improvements are expected to bring the computational scaling closer to that of TDDFT analytic gradients and will allow us to assess the accuracy of BSE@$GW$ across a wider range of molecular systems.
In this context, the manifestation of discontinuities of $GW$ quasiparticle energies on excited-state properties will be investigated. \cite{veril2018unphysical,loos2020pros,monino2022unphysical} 
Furthermore, the possibility of determining analytic properties for double excitations within the dynamical BSE will be explored.

The extension to second-order derivatives for determining vibrational harmonic frequencies that are directly experimentally accessible is also of interest.
Lastly, this work paves the way toward the derivation and implementation of non-adiabatic couplings within the BSE framework, which would allow for the investigation of non-adiabatic effects in molecular dynamics simulations based on the Bethe--Salpeter equation formalism. 

\section*{Supporting Information}
The Supporting Information (SI) contains additional details regarding drUCCD, the BSE@$GW$ Lagrangian, validation of the nuclear gradients, and a comparison of BSE@$GW$@BHLYP and TDDFT@BHLYP transition energies.

\section*{Acknowledgement}

J.~T.~acknowledges funding from the Fonds der Chemischen Industrie (FCI) via a Liebig fellowship and support by the Cluster of Excellence ``CUI: Advanced Imaging of Matter'' of the Deutsche Forschungsgemeinschaft (DFG) (EXC 2056, funding ID 390715994).
M.-P.~K.~and P.-F.~L.~thank the European Research Council (ERC) under the European Union's Horizon 2020 research and innovation programme (Grant agreement No.~863481) for funding.
For this work the HPC-cluster Hummel-2 at University of Hamburg was used. The cluster was funded by Deutsche Forschungsgemeinschaft (DFG, German Research Foundation) - 498394658.
% references
%

\end{document}